\def\din-A4{\setlength{\oddsidemargin}{1.2\oddsidemargin}
  \addtolength{\oddsidemargin}{1.2\evensidemargin}
  \setlength{\topmargin}{1.2\topmargin}
  \addtolength{\topmargin}{-15.4mm}
  \setlength{\textwidth}{1.2\textwidth}
  \setlength{\textheight}{1.2\textheight}
  \setlength{\evensidemargin}{0.5\oddsidemargin}
  \setlength{\oddsidemargin}{\evensidemargin}
  \addtolength{\oddsidemargin}{-18.7mm}
  \addtolength{\evensidemargin}{-18.7mm}
  \setlength{\tabcolsep}{1.2\tabcolsep} }

\documentclass[12pt]{article}
\usepackage{amssymb}
\usepackage{latexsym}

\newtheorem{theorem}{Symmetric Space Theorem}
  
\newtheorem{lemma}{Lemma} 
\newtheorem{cor}{Corrolary} 

\def\C{\mathbb{C}}
\def\R{{\mathbb R}}
\def\Z{{\mathbb Z}}

\def\Vir{\mbox{Vir}}
\def\g{\mathfrak{g}}
\def\p{\mathfrak{p}}   
\def\q{\mathfrak{q}}      
\def\u{\mathfrak{u}} 
\def\Loop{{\cal L}}    
\def\Lg{\Loop(\g)}   
\def\Lgt{\tilde\Lg}   
\def\Lu{\Loop(\u)}  
\def\Lut{\tilde\Lu}
\def\id{\mbox{id}}
\def\dg{\dim\g}
\def\qed{\hfill\qquad\mbox{$\Box$}}
\def\proof{{\em Proof: }}
\def\be{\begin{equation}} \def\rf#1{(\ref{#1})}
\def\br{\begin{eqnarray}} \def\er{\end{eqnarray}}
\def\be{\begin{equation}} \def\ee{\end{equation}} \def\rf#1{(\ref{#1})}
 \def\nn{\nonumber}

\begin{document}
\title{Algebraic Proof of the Symmetric Space Theorem}
\author{{\it C.\ Daboul} 
\\Mathematisches Seminar der Universit\"at Hamburg, \\Bundesstr.55,
D-20146 Hamburg, Germany} 
\maketitle

\begin{abstract} I give a relatively elementary proof of the symmetric
 space theorem, due to Goddard,
 Nahm and Olive \cite{GNO}.  Unlike their original proof, which
involves the quark-model construction,  I only use elementary algebraic
techniques. \end{abstract}

\section{Introduction}
\label{Intro}
In 1985 Goddard, Nahm and Olive \cite{GNO} proved an interesting
theorem, which has been called the {\em symmetric space theorem}. It
relates the vanishing of certain coset Virasoro
algebras to the existence of symmetric spaces.
 This has further interesting
mathematical consequences in  representation  theory of affine
Kac-Moody algebras, since the vanishing of the coset Virasoro algebra
is the
 condition for finite  reducibility of representations
of affine algebras restricted to certain affine subalgebras.

The original proof by the above authors involved the use of physical
concepts, such as quarks. The purpose of the present paper is to
state and prove the symmetric space theorem, by using purely algebraic
concepts. In particular, I shall carry out the proof following
 an idea used by Witt \cite{Witt}, thereby using only
elementary algebraic transformations.

In section \ref{DefNo} I introduce relevant definitions and notations.
  In section \ref{Theo} I first prove three auxiliary lemmas and two
corollaries, then state and prove the theorem.

\section{Definitions and Notations}  
\label{DefNo}
\subsection{Affine Algebras} 
The {\em affine Kac-Moody algebra}, affine algebra for short, 
\be\label{affalg}
\tilde
\Loop (\g) = (\C [t,t^{-1}]\otimes_\C \g) \oplus\C {\cal K} \ee
associated to an  underlying
simple Lie algebra $\g$ is defined by the following commutation
relations: 
\be\label{affcomm} 
[t^m\otimes x,t^n\otimes y]=t^{m+n}\otimes [x,y]+m\;
\delta_{m,-n}(x|y) {\cal K} 
\end{equation} where ${\cal K}$ spans the
center
 of $\Lgt$ and is called the {\em central term},
and $(.|.)$ is  the {\em normalized invariant form} on $\g$.  The term
invariant means that 
\be\label{inv} ([x,y]|z)=(x|[y,z])\qquad
\forall\,x,y,z\in\g. \end{equation}
It is well known (see for
example \cite{Hum}), that two invariant symmetric bilinear forms on a
simple Lie algebra differ only by a scalar factor.  The normalized
invariant form, which will be used henceforth, is defined such that
the  highest  root has length $\sqrt{2}$. 

The affine algebra for an underlying  abelian algebra $\g$
can be defined similarly, by \rf{affalg} and
\rf{affcomm}.
However, for abelian $\g$ any nondegenerate 
symmetric form $(.|.)$ is invariant. In contrast to the case of a
 simple algebra, such a form cannot be further determined by algebraic
constraints derived from the Lie algebra structure.
In this article abelian algebras will almost always appear as
subalgebras of the simple algebra $so(n)$. Therefore, there will be a
natural choice of $(.|.)$, namely the restriction of the normalized
invariant form of $so(n)$ to $\g$.
In the definitions, where an abelian algebra is not given as a
subalgebra of $so(n)$, $(.|.)$ may be any fixed 
nondegenerate  symmetric form.

When the central element ${\cal K}$ of the affine algebra $\Lgt$
acts as a scalar $k\,\id$ in a representation, then $k$ is called
the {\em level of this representation}.

   For a reductive Lie algebra 
\be \label{red}
\u=\u_0\oplus\u_1\oplus\dots\oplus\u_{S} \end{equation}
 with center $\u_0$ and simple ideals $\u_1,\dots ,\u_S$, the affine
algebra $\Lut$ is the direct sum of the affine algebras associated to
the ideals $\u_0,\dots,\u_S$:
 \be  \Lut =\bigoplus_{s=0}^{S}\;\tilde \Loop (\u_s)=
\bigoplus_{s=0}^{S}\; ((\C [t,t^{-1}]\otimes\u_s)\, \oplus\C
{\cal K}_{s}). \end{equation}

The affine algebra associated to a finite-dimensional reductive Lie
algebra $\u$ is thus a $S$-dimensional central extension of the {\em
loop algebra} $\C [t,t^{-1}]\otimes\u$ {\em associated to
$\u$}.
(In the case of an underlying semi-simple Lie algebra 
the
affine algebra is indeed the universal central extension of the loop
algebra.)
 The loop algebra can be identified with the Lie algebra of
polynomial maps of $S^1$ into $\u$.

\subsection{The Virasoro Algebra and the Sugawara Construction}
The {\em Virasoro algebra} 
$$\Vir=\bigoplus_{n\in\Z} \C d_n
\oplus\C c$$ 
is defined by the
following commutation relations: 
\begin{eqnarray}[c,d_n]&=&0\nonumber\\
\label{vir}
[d_m,d_n]&=&(m-n)d_{m+n}+\delta_{m,-n}\frac{m^3-m}{12}\,c\qquad\qquad
\forall\,m,n \in\Z.
\end{eqnarray}
Let $\g$ be a simple Lie algebra.
Let $\{u_i\}$ and
$\{u^i\}$ be dual bases of $\g$.
The {\em quadratic Casimir operator} is the following element of the
universal enveloping algebra $\mathfrak U(\g)$ of $\g$:
  \be\label{casi} 
\Omega = \sum_{i=1}^{\dim\g}\,u_i\,u^i.
\ee 
$\Omega$ commutes with any element of $\g$, therefore, by Schur lemma
it acts as a scalar, denoted by $\omega_\rho$, on any irreducible
representation $\rho$ of $\g$.
Given a representation of the affine algebra $\Lgt$ of level
$k\ne-\frac{1}{2}\,\omega_{ad}$,
where $\omega_{ad}$ denotes the scalar value of the quadratic Casimir
operator in the adjoint
representation, one can construct a representation of
$\Vir$ by using the following {\em Sugawara operators} 
(cf.\ \cite{Kac}), 
where the notation $x^{(n)}\equiv t^n\otimes x$ is used:
\begin{eqnarray}
\nonumber
L_0&:=&\frac{1}{(2k+\omega_{ad})}\,\left(
\sum_{i}u_iu^i+2\sum_{n=1}^{\infty}\sum_{i}u_i^{(-n)}u^{i(n)}\right)\\
\label{suga}
L_n&:=&\frac{1}{(2k+\omega_{ad})}\,\sum_{m\in\Z}\sum_{i}u_i^{(-m)}u^{i(m+n)}
\qquad\qquad\forall\,{n\in\Z\setminus\{0\}}. \end{eqnarray}
The operators $L_n$ obey the commutation relations \rf{vir}, where 
the  central element  of the Virasoro algebra takes the scalar value
 \begin{equation} \label{cSug}
 c = \frac{2k\dg}{2k+\omega_{ad}}. \end{equation}  For an underlying
abelian algebra the same equations hold with  $\omega_{ad} = 0$. In
this case \rf{cSug} yields $c = \dg$.
The Sugawara
operators $L_m$ and the elements $x^{(n)}$ of the affine algebra
obey the following commutation relations: 
\be
\label{virkm}
[L_m,x^{(n)}]=-n x^{(m+n)}\qquad\forall\,{m,n\in\Z,\,x\in\g}.
\end{equation}
For a reductive Lie algebra as in \rf{red}, the 
Sugawara operators corresponding to $\Lut$ in a representation of
levels $k_s\ne-\frac{1}{2}\,\omega_{ad_s}$ are given by 
 $$L_n^{\u}:=\sum_{s=0}^S L_n^{\u_{s}},$$ where $L_n^{\u_{s}}$
is the Sugawara operator for the individual ideal $\u_s$.
The $L_n^{\u_{s}}$ provide a representation of the Virasoro algebra
 with central 
value
\be\label{cu1}
c_{\u}=\sum_{s=0}^S \frac{2k_s\dim
\u_s}{2k_s+\omega_{ad_s}}.
\ee 
Again, the $L_n^{\u}$ satisfy a commutation relation similar
to \rf{virkm}, where $x\in\u$. 

\subsection{The Coset Construction}  
\label{cc}
Let $U$ be a compact Lie subgroup
of the orthogonal group $SO(n)$. Then the complexified Lie algebra of
$U$ is a reductive algebra $\u$ as in \rf{red}, which is
a subalgebra of the orthogonal
 algebra $so(n)$.
 
The inclusion of $\u$ in $so(n)$ gives rise to a homomorphism  of the
associated affine algebras:  
\be \label{homo} \Lut =
\bigoplus_{s=0}^{S}\; ((\C [t,t^{-1}]\otimes\u_s)\, \oplus\C
{\cal K}_{s}) \;\to\; \tilde \Loop (so(n)) = (\C [t,t^{-1}]\otimes
so(n)) \oplus\C {\cal K}. \end{equation} The obvious inclusion homomorphism
of
the loop algebras is lifted  consistently to a homomorphism of the
affine algebras in \rf{homo} by letting \be \label{Ks} {\cal K}_s\; 
\mapsto
j_s\, {\cal K} \qquad \mbox{for } s\, =\, 0,\dots, S,  \end{equation}
where for the simple ideals $\u_s,\;s\ge 1$, the factor
$j_s$ is the {\em Dynkin index}, which is defined 
as the ratio of the normalized invariant form $(.|.)_{so(n)}$ on
$so(n)$ (restricted to $\u_s$) and the normalized invariant form 
$(.|.)_s$ on $\u_s$   i.e.\ 
\be\label{js}
(x|y)_{so(n)} = j_s\,(x|y)_s  \quad\forall\,{x,y\in\u_s}.
\ee 
For $s=0$ we choose $(.|.)_0$ to be the restriction of 
$(.|.)_{so(n)}$ to $\u_0$. Then, by letting $j_0=1$, equation \rf{js}
holds also for $s=0$.

In this situation the Sugawara construction can be applied to both
$so(n)$ and $\u$ to obtain Sugawara operators $L_n^{so(n)}$ 
and $L_n^{\u}$ which are different in general and form
representations of $\Vir$. We can calculate their central values
by using \rf{cSug} and \rf{cu1}. We get
\be c_{so(n)}= \frac{k_{so(n)}\, n\,(n-1)}{2k_{so(n)}+2n-4}
          \label{cso(n)}
\end{equation}
and 
\be
c_{\u}=
\sum_{s=0}^S \frac{2k_{so(n)}\, j_s\dim \u_s}
                        {2k_{so(n)}\, j_s+\omega_{ad_s}},
\label{cu}
\end{equation}
respectively.
In \rf{cso(n)} I substituted $\dim(so(n))=n(n-1)/2$ and
   \begin{equation} \label{OadVal}
 \omega_{ad_{so(n)}}= 2n-4. \end{equation} (The expression \rf{OadVal}
can be computed, for example, by using the relation 
$\omega_{ad}=2\check h$,
 where $\check h$ is the {\em dual Coxeter number}. A table of dual
Coxeter numbers can be found in \cite{Kac}.)
In \rf{cu} I used $k_s=j_s\, k_{so(n)}$, which follows from
\rf{Ks}.

By \rf{virkm} we have
\be
[L_l^{so(n)},x^{(m)}]=-m x^{(m+l)}\qquad\forall\,{l,m\in\Z,\,x\in so(n)}.
\end{equation}
\be
[L_l^{\u},x^{(m)}]=-m x^{(m+l)}\qquad\forall\,{l,m\in\Z,\,x\in\u}.
\end{equation}
Therefore, the difference operators
$$K_l := L_l^{so(n)}-L_l^{\u}$$
commute with each $x^{(m)}\in\Lut$ individually
\be
[K_l,x^{(m)}]=0\qquad\forall\,{l,m\in\Z,\,x\in\u},
\end{equation}
and consequently, by \rf{suga}, with the corresponding Sugawara
operator
\be
[K_l,L_m^{\u}]=0\qquad\forall\,{l,m\in\Z}.
\end{equation}
It follows, that
\be
[K_l,K_m]=[L_l^{so(n)},L_m^{so(n)}]-[L_l^{\u},L_m^{\u}]
\qquad\forall\,{l,m\in\Z}.
\end{equation}
We deduce that the $K_m$, like the $L_m^{so(n)}$ and the $L_m^{\u}$,
define a representation of $\Vir$, whose central charge is equal 
to the difference of the Sugawara values 
\begin{eqnarray} c_K &=&
c_{so(n)}-c_{\u}\nonumber\\
  &=& \frac{k_{so(n)}\, n\,(n-1)}{2k_{so(n)}+2n-4}-
          \sum_{s=0}^S \frac{2k_{so(n)}\, j_s\dim \u_s}
                        {2k_{so(n)}\, j_s+\omega_{ad_s}}. \label{ccos}
\end{eqnarray} 
The above construction of representations of the Virasoro algebra is
called the {\em coset construction}. 
It was introduced by Goddard and Olive in \cite{GO1}.
It can be applied similarly for 
any simple
(or even reductive) Lie algebra $\g$  instead of $so(n)$  and a
reductive subalgebra $\u\subseteq\g$, but in this article I shall only
 consider the case $\g=so(n)$.
 
An important property of all the above constructions is, that they 
preserve the unitarity of the involved representations i.e.\ a
unitary representation of the affine algebra $\tilde\Loop(so(n))$, 
induces a unitary representation of its subalgebra
$\Lut$. Then  by  the Sugawara construction applied to the
unitary representations of $\Lut$ and $\tilde\Loop (so(n))$ we get two
different unitary representations of $\Vir$. Finally,  the resulting
coset representation of $\Vir$ is again unitary.

The Virasoro algebra has no 
nontrivial unitary representations with zero central charge, so that 
the coset Virasoro algebra vanishes iff 
$c_K=0$.
 Furthermore, as was first shown in \cite{GO1}, the
representation of $\tilde\Loop (\u)$ induced by a unitary highest
weight representation of $\tilde\Loop (so(n))$ is finitely reducible,
if and only if the coset Virasoro algebra vanishes.

It can be shown that $c_K=0$ is only possible for level $1$ 
representations of $\tilde\Loop (so(n))$, (see e.g.\ \cite{dab}). In
this case \rf{ccos} reduces to
 \be \label{form1} c_K = \frac{n}{2}- \sum_{s=0}^S\frac{2j_{s}\dim
\u_{s}}{2j_{s}+\omega_{ad_s}}\qquad\mbox{for } k_{so(n)}=1. 
\end{equation}

\subsection{Indices of Representations} 

The {\em index $\kappa_{\rho}$ of a representation} $\rho$ of a simple
 Lie algebra is defined as the ratio between the 
trace form of the  representation and the normalized invariant form on
the algebra, i.e.\ 
\be\label{kapdef}
Tr(\rho(x)\rho(y)) = \kappa_{\rho}\,(x|y).
\ee   

For a simple Lie algebra $\g$ or an abelian subalgebra of a simple 
 Lie algebra in an $n$-dimensional representation $\rho$ of $\g$, on
which the Casimir operator is a scalar multiple of the identity, 
 $\rho(\Omega)=\omega_{\rho}\,\id_n$
one gets by using \rf{casi}
and \rf{kapdef}:
\be \label{kapdg}
\kappa_{\rho}\dim\g=\omega_{\rho}n. \end{equation}  
In particular, if $\rho$ is the adjoint
representation of a simple
 Lie algebra, then \rf{kapdg} yields for its index
\be \label{kappa_ad}
\kappa_{ad}=\omega_{ad}.
\ee
This equation also holds
(trivially)  for an abelian Lie algebra, since in this case 
$\kappa_{ad}=\omega_{ad}=0$.

It can be shown, that the index of the natural
representation of $so(n)$ (i.e.\ the representation of $so(n)$ by
antisymmetric $n\times n$-matrices) is $2$ (see e.g.\ \cite{dab}). Let
$\u$ be a reductive subalgebra of $so(n)$ as in subsection \ref{cc}.
Then the index $\kappa_{\rho_s}$ of the representation $\rho_s$ of the
ideal $\u_s$ obtained by restricting the natural representation of
$so(n)$ to $\u_s$ is determined by 
$$ Tr(\rho_s(x)\rho_s(y))=2\, (x|y)_{so(n)}= 2\,
j_s\, (x|y)_{s}\quad\forall\, 
 x,y\in\u_s$$ i.e.\  
\be\label{kj} \kappa_{\rho_s}= 2\,j_s.\ee 
Note, that with our choice of
$(.|.)_0$ the definition of the
index makes sense also for the abelian subalgebra $\u_0$ in
the representation $\rho_0$ and we get $\kappa_0=2$.

\subsection{Infinitesimal Symmetric Spaces}  Let $\g$ be a semi-simple
Lie algebra and let $\sigma$ be an {\em involution} of $\g$, i.e.\ an 
automorphism of $\g$ of order $2$:  $\sigma^2=\id$. 
Let  
$$\g=\g_0\oplus\g_1$$ be the decompositon of $\g$ into eigenspaces of
$\sigma$ $$ \g_0=\{x\in \g\mid\sigma(x)=x\},\quad\g_1=\{x\in
\g\mid\sigma(x)=-x\},$$ then $\g/\g_0\cong\g_1$ is called an {\em
infinitesimal symmetric space}.

Under the above conditions the following relations hold
\be\label{Z2grad} [\g_0,\g_0]\subseteq\g_0,\quad
[\g_0,\g_1]\subseteq\g_1,\quad [\g_1,\g_1]\subseteq\g_0. \end{equation}
This means that the Lie algebra $\g$ is $\Z/2\Z$-graded. On the other hand,
given a $\Z/2\Z$-gradation on $\g$ (i.e.\ a decomposition
$\g=\g_0\oplus\g_1$ such that \rf{Z2grad} holds), then an involution
$\sigma$  of $\g$ is defined by letting $\sigma(x)=x$ for
all $x\in\g_0$ and $\sigma(x)=-x$ for all $x\in\g_1$. Thus the
infinitesimal symmetric spaces can equivalently be  defined by
\rf{Z2grad}.

Since the Killing form is invariant under automorphisms, we have
\br
\nonumber
Tr(ad(x) ad(y))&=&
Tr(ad(\sigma(x)) ad(\sigma(y)))\\
\label{invf}&=&-Tr(ad(x) ad(y))
\qquad\forall\,x\in\g_0,\,y\in\g_1.
\er
Thus $\g_0$ and $\g_1$ are orthogonal with respect to the Killing form
on $\g$ and the restriction of the Killing form to $\g_0$ respectively
to $\g_1$  is nondegenerate.\\

The list of all symmetric spaces is due to E.\ Cartan. A derivation
 based on Kac' classification of finite order automorphisms of
semisimple Lie algebras can be found in \cite{Helg}, cf.\ also
\cite{GNO}. 
 
\subsection{The underlying real representation of an orthogonal
representation}
A complex  representation of a compact Lie group $U$ is called
{\em orthogonal} if there exists a nondegenerate symmetric bilinear
form $(.|.)$ on the representation space $\p$, such that
\be (g(x)|g(y))=(x|y)\qquad\forall\,g\in U,\,x,y\in \p,\ee
i.e.\, the form is {\em invariant} under $U$.
By choosing an orthonormal base of $\p$ with respect to $(.|.)$ we get a
matrix representation of $U$ by orthogonal matrices. If $U$ is
connected and the orthogonal representation is faithfull this gives
 an embedding of $U$ 
in $SO(n)$.

On the Lie algebra level the invariance property is the following
\be
(u(x)|y)=-(x|u(y))\qquad\forall\,u\in \u,\,x,y\in \p,
\ee 
for example \rf{inv} is equivalent to this equation in the special
 case
of the adjoint representation. 
Since the Killing form is invariant under the adjoint action and 
since it is
 nondegenerate for a semisimple Lie algebra, the adjoint
representation of a semisimple Lie algebra $\g$ is an
orthogonal  representation.

In the case of a $\Z/2\Z$-graded Lie algebra, the representation
of $\g_0$ on $\g_1$ is an orthogonal representation, the restriction
of the Killing form to $\g_1$ beeing an invariant form. 

It is well known (see e.g.\ \cite[Prop.(6.4)]{BtD}), that every
complex orthogonal representation of a compact Lie group $U$ 
is the complexification of a real representation of $U$. This means,
that there exists a real subspace $\p^{\R}$ of $\p$, invariant under the
group action ( $U(\p^{\R})\subseteq\p^{\R}$ ), such that 
$$\p=\p^{\R}\oplus i\,\p^{\R}$$
as a real vector space, (i.e.\ each element of $p\in\p$ can be uniquely
decomposed as $p=x+iy$ with $x,y\in \p^{\R}$) and such that
$$u(x+iy)=u(x)+i\,u(y)\quad\forall\,{u\in U,\, x,y\in \p^{\R}}.$$
Although $\p^{\R}$ is in general not uniquely determined as a real
subspace  of $\p$, its isomorphic type as a real $U$-module is uniquely
determined by the isomorphic type of the complex  $U$-module $\p$,
since $\p$ is isomorphic  to $\p^{\R}\oplus\p^{\R}$ as a real
$U$-module.  

\section{The Symmetric Space Theorem}
\label{Theo}
The following assumptions and notations shall be valid for the whole
section:
{\em
Let $U$ be a compact Lie group
and let the reductive Lie algebra
$\u=\u_0\oplus\u_1\oplus\dots\oplus\u_{S}$ 
 with center $\u_0$ and simple ideals $\u_1,\dots ,\u_S$
be the complexification of the Lie algebra of $U$.
 Let
$\{u^i\}_{i=1,\dots,\dim\u}$ be a base for $\u$ consisting of
antihermitian elements, which is the union of bases $\{u^i\}_{i\in
I_{s}}$ for the ideals $\u_s$ such that \mbox{$(u^i|u^j)=-\delta_{ij}$}
for $i,j\in I_{s}$. 
(This is possible since on the real
subalgebra of antihermitian elements the
 normalized invariant form is negative definite.)  
 For $i\in\{1,\dots,\dim\u\}$ let $s(i)$ 
denote the index of the ideal, which contains the element $u^i$,
$u^i\in\u_{s(i)}$,
such that $0\le s(i)\le S$.}

Before stating the theorem I shall prove three lemmas, which will be
needed for its proof, but can also be useful by themselves. 

\begin{lemma} \label{lsw} 
The following equation
holds in the universal enveloping algebra $\mathfrak U(\u)$  and thus in
any representation of $\u$
\be\label{sw} 2\sum_{i\in I_{s}}
u^iu^ju^i = 
-2\Omega_s  u^j
+\delta_{s(j),s}\omega_{ad_{s}}u^j \qquad\mbox{for } 0\le s\le S,
\quad 1\le j\le \dim\u,
\end{equation} 
where $\Omega_s $ denotes the
quadratic Casimir operator of the ideal $\u_s$.
\end{lemma}
\proof The
quadratic Casimir operator of $\u_s$ can be written as
$$\Omega_{s}=-\sum_{i\in I_{s}}(u^i)^2.$$ 
Applying $ad_{\u}(\Omega_{s})$ to $u^j$ gives $\omega_{ad_{s}} u^j$,
if $u^j\in\u_s$ and $0$ otherwise, since $[u^i,u^j]=0$ for $s(i)\not
= s(j)$. Therefore,
\begin{eqnarray*}
ad_{\u}(\Omega_{s})\,u^j=\delta_{s(j)s}\,\omega_{ad_{s}} u^j &=&
-\sum_{i\in I_{s}}ad(u^i)^2 (u^j)
=-\sum_{i\in I_{s}} [u^i\,[u^i,u^j\,]\,]\\
  &=&-\sum_{i\in
I_{s}}\left( u^i(u^iu^j-u^ju^i)-(u^iu^j-u^ju^i)u^i\right)
\\  
  &=&-\sum_{i\in I_{s}}
\left((u^i)^2u^j-u^iu^ju^i-u^iu^ju^i+u^j(u^i)^2\right)
\\  
&=& 2\Omega_s  u^j
 +2\sum_{i\in I_{s}}
u^iu^ju^i,
\end{eqnarray*}
where we used, that $\Omega_s $ commutes with $u^j$.\qed

Applying \rf{sw} again to
$u^j$ and summing over $j$ we immediately get 
\begin{cor} \label{lbeforetr} 
The sum of the elements $(u^i\,u^j)^2$ lies in the center of
$\mathfrak U(\u)$ and is given in terms of the quadratic Casimir operators
as follows:
\be\label{beforetr} 2\sum_{i\in I_{s}}\sum_{j\in I_{t}}
(u^i\,u^j)^2 =\left( 
2\Omega_{t}
-\delta_{st}\omega_{ad_{s}}\right)\Omega_s \quad\mbox{for }
0\le s,t\le S.
\end{equation} \end{cor}

 By taking the
trace of  equation \rf{beforetr} we immediately get the following
corollary, which will be used in the proof of the next lemma : 
\begin{cor} \label{ltr} 
In any $n$-dimensional matrix representation of $\u$, such that the
operators $u^i$ are represented by matrices $M^i$ and such
that the quadratic Casimir operator $\Omega_s$ of the ideal $\u_s$
acts as a scalar $\omega_{{s}}\id_n$ for $s=0,\dots,S$, the
following identity holds:
 \be
\label{tr}  2\sum_{i\in I_{s}}\sum_{j\in I_{t}}
Tr\left((M^iM^j)^2\right) = \left( 
2\omega_{{t}}
-\delta_{st}\omega_{ad_{s}}\right)\omega_{{s}}\,n\,
\quad\mbox{for }
0\le s,t\le S. \ee
\end{cor}

The following further assumptions and notations shall again be valid 
in the sequel:\\

{\em
Let $\p$ be a faithful
$n$-dimensional orthogonal complex representation of the compact
group $U$. This defines an embedding of $U$ into $SO(n)$ and we shall
thus view $U$ as a subgroup of $SO(n)$.
 Let $\{p_{\alpha}\}_{\alpha=1,\dots, n}$ be an
orthonormal base of $\p^{\R}$. Let the application of the operators
$u^i$ on the basis $p_{\alpha}$ be described by the real
antisymmetric matrices $M^i$, as follows 
  \be\label{Mdef}  u^i(p_\alpha)=
\sum_{\gamma=1}^n M_{\gamma\alpha}^i \, p_\gamma~, \end{equation}
 Furthermore let
$y_s$ denote the index of the representation of $\u_s$ on
$\u\oplus\p$.}\\

To evaluate $y_s$ we note that (a) the index of a direct sum of
representations is the sum of the indices of the direct summands and (b)
the index of the adjoint action of $\u_{s}$ on $\u$ is $\omega_{ad_{s}}
$, because of \rf{kappa_ad} and since  the adjoint action of $\u_{s}$
on the sum of the $\u_{t}$ with $t\not = s$ is trivial.  Denoting the
index of the representation of the ideal $\u_s$ on $\p$ by 
$\kappa_{{s}}$ (instead of $\kappa_{\rho_s}$),
we get
\be \label{ymu}
y_{s}=\omega_{ad_{s}}+\kappa_{{s}}=
\omega_{ad_{s}}+2 j_{s}. \end{equation} 
It follows, that $y_{s}> 0$ for $s=0,\dots,S$.

{\em We assume,
 that the underlying real representation $\p^{\R}$ of $\p$ is
irreducible.} This holds for example if $\p$ is itself
irreducible, but $\p$ can also decompose as
$\p\cong\q\oplus\q^*$, where $\q$ is some irreducible complex
representation, such that $\q\not\cong\q^*$. 
(The theorem also
holds when $\p^{\R}$ is reducible,  and it can be derived from
 the irreducible case (see e.g.\ \cite{dab}).)

It can be shown that under the assumption, that
$\p^{\R}$ is irreducible as a $U$-module, the
Casimir operators $\Omega_s $ act as real positive scalars on $\p$,
although $\p$ is not  irreducible as a module over $\u_s$,
\be\label{scalarCasimirs}
\rho(\Omega_{s})=\omega_{{s}}\id_n\qquad\mbox{with
}\omega_{{s}}>0. \ee
Therefore, \rf{kapdg} and \rf{tr} hold
 in this situation. 

\begin{lemma} \label{lJ}
Let V be a level one
representation space of the affine algebra $\tilde\Loop (so(n))$.
Let $c_\u$ denote the central element of the Virasoro algebra
constructed from $\Lut$ by the Sugawara construction on V and let $c_K$
denote the central element of the coset Virasoro algebra on V. 
Furthermore,
let 
\be\label{jdef}
 J_{\alpha\beta\gamma\delta}:= \sum_{s=0}^S \frac {1}{y_{s}}
\sum_{i\in I_s} \left( M^i_{\gamma\beta}M^i_{\delta\alpha}
+M^i_{\beta\alpha}M^i_{\delta\gamma}+M^i_{\alpha\gamma}M^i_{\delta\beta}
\right)~, \ee where $M$ is defined in \rf{Mdef}, then the following
identity holds 
\be\sum_{\alpha\beta\gamma\delta} 
\left(J_{\alpha\beta\gamma\delta}\right)^2 = \frac{6}{n}\,c_{\u}\,c_K.
\label{g3l} \ee \end{lemma} 
\proof First let us bring $c_K$ and
$c_{\u}$ in the form, that is most adequate to the following
calculations. 
Substituting \rf{kj} 
in \rf{form1} we get  
\be \label{formb} c_K ~=~\frac{n}{2}-
\sum_{s=0}^S\frac{\kappa_{s}\dim \u_{s}}{\kappa_{s}+\omega_{ad_{s}}}~
\stackrel{\rf{kapdg}\rf{ymu}}=~
\frac{n}{2}-\sum_{s=0}^S\frac{n\omega_{{s}}}{y_{s}}.   \ee For
$c_{\u}$ alone we have \be
c_{\u}=\sum_{s=0}^S\frac{n\omega_{{s}}}{y_{s}}>0. \ee

The sums \be J^i_{\alpha\beta\gamma\delta}:=
M^i_{\gamma\beta}M^i_{\delta\alpha}
+M^i_{\beta\alpha}M^i_{\delta\gamma}+M^i_{\alpha\gamma}M^i_{\delta\beta}~
, \ee
 are invariant under cyclic permutation of the indices
$\alpha,\beta,\gamma\,$:
$J^i_{\alpha\beta\gamma\delta}=J^i_{\gamma\alpha\beta\delta}=
J^i_{\beta\gamma\alpha\delta}~,$ so that
\begin{eqnarray} \frac{1}{3}\sum_{\alpha,\beta,\gamma,\delta}
J^i_{\alpha\beta\gamma\delta}J^j_{\alpha\beta\gamma\delta}&=& 
\sum_{\alpha,\beta,\gamma,\delta} J^i_{\alpha\beta\gamma\delta}\;
M^j_{\gamma\beta}M^j_{\delta\alpha}\nn\\
&=& \sum_{\alpha,\beta,\gamma,\delta}
(M^i_{\gamma\beta}M^i_{\delta\alpha}+M^i_{\beta\alpha}M^i_{\delta\gamma}
+M^i_{\alpha\gamma}M^i_{\delta\beta})
M^j_{\gamma\beta}M^j_{\delta\alpha}\nonumber\\
&=&(Tr(M^iM^j))^2
-2Tr\left((M^iM^j)^2\right)\nonumber\\
&=&\delta_{ij}\,\kappa^2_{s(i)}
-2 Tr\left((M^iM^j)^2\right)
~,\label{ij} \end{eqnarray} 
where $\kappa_{s}$ is the
index of the representation of $\u_s$ on $\p$. For the third 
equality I used the antisymmetry of the matrices $M^i$. 
Using \rf{ij} we get 
\begin{eqnarray*} 
\frac{1}{3}\sum_{\alpha,\beta,\gamma,\delta}
\left(J_{\alpha\beta\gamma\delta}\right)^2
 \!\!\!\!\!\!\!\!\! &=& \!\!\!\!\!\!
\frac 1{3}\sum_{s,t=0}^S\frac{1}{y_{s}y_{t}}\sum_{i\in I_{s},j\in
I_{t}}\sum_{\alpha,\beta,\gamma,\delta}
J^i_{\alpha\beta\gamma\delta}J^j_{\alpha\beta\gamma\delta}\nn\\ 
 &\stackrel{\rf{ij}}=&\label{ji2}
\sum_{s=0}^S\frac{\kappa_{s}^2\dim\u_{s}}{y_{s}^2}
-\sum_{s,t=0}^S\frac{2}{y_{s}y_{t}}\sum_{i\in I_{s},j\in I_{t}}
Tr\left((M^iM^j)^2\right)\\
 &\stackrel{\rf{kapdg}\rf{tr}}=&
\sum_{s=0}^S\left( \frac{\kappa_{s}\omega_s n}{y_{s}^2}
-\sum_{t=0}^S\frac{\left(2\omega_{{t}}
-\delta_{st}\omega_{ad_{s}}\right)\,
\omega_{{s}}\,n}{y_{s} y_{t}}\right)
 \nonumber\\
&\stackrel{\rf{ymu}}=&
\sum_{s=0}^S \frac{\omega_s n}{y_{s}}
\left( \frac{\kappa_{s}+\omega_{ad_{s}}}{y_{s}}
-\sum_{t=0}^S\frac{2\omega_{{t}}}{y_{t}}
\right)\\
&=&\left(\sum_{s=0}^S\frac{\omega_{{s}}n}{y_{s}}\right)
\left(1-\sum_{t=0}^S\frac{2\omega_{{t}}}{y_{t}}\right) \\
&\stackrel{\rf{formb}}
=& c_{\u}\cdot\frac{2\,c_K}{n}
\qquad\qquad\qquad
\qquad\qquad\qquad\mbox{\qed} \end{eqnarray*} 

The following lemma was shown in \cite{GNO}:

\begin{lemma}\label{lprod}
Suppose that $\g\equiv\u\oplus\p$ forms a $\Z/2\Z$-graded algebra, as in
\rf{Z2grad}  with $\g_0=\u$
and $\g_1=\p$, which is related  to the given orthogonal 
$\u$-module-structure on $\u\oplus\p$ as follows: 
\begin{itemize}

\item{ The Killing form of  $\g$ coincides on $\p\times\p$ with the 
given orthogonal inner product, i.e.
$Tr(ad_{\g}(p_\alpha)ad_{\g}(p_\beta))=\delta_{\alpha,\beta}\quad
\forall\,1\le\alpha,\beta\le n$. }  

\item{ The restriction  of the adjoint representation of $\g$ to $\u$
coincides  with the given $\u$-module structure of $\u\oplus\p$, i.e.}
\br[u^i,u^j+p_\alpha]_{\g}&=&[u^i,u^j]_{\u}+u^i(p_\alpha)  \\
&=&
[u^i,u^j]_{\u}+\sum_{\gamma} M^i_{\gamma\alpha} p_\gamma~, \quad
 1\le i,j \le \dim\u \;,\; 1\le\alpha \le n \nn\er
where  $[.\,,.]_{\g}$ and
$[.\,,.]_{\u}$ denote the Lie products on $\g$ and $\u$,
respectively.
\end{itemize}

 Then the Lie product on $\p$ is given by 
\be  \label{struktur}
[p_{\alpha},p_{\beta}]_{\g}=\sum_{i=1}^{\dim\u}
\frac{1}{y_{s(i)}}M^i_{\alpha\beta} u^i.  \end{equation}  \end{lemma}
\proof The  invariance \rf{inv} of the Killing form means that
\be\label{Formen}
Tr(ad(u^i)\,ad([p_{\alpha},p_{\beta}]))=Tr(ad([u^i,p_{\alpha}])\,ad(
p_{\beta})).
\end{equation} Since $[\g_1,\g_1]\subseteq \g_0$ by assumption, we must
have  $[p_{\alpha},p_{\beta}]=\sum_{j=1}^{\dim\u}X_{\alpha\beta}^{j}
u^j$, where the $X_{\alpha\beta}^{j}$ are some constants, such that
$X_{\alpha\beta}^{j}=-X_{\beta\alpha}^{j}$. These constants
can be determined by equating the l.h.s.\ of \rf{Formen}
$$\sum_{j=1}^{\dim\u}X_{\alpha\beta}^{j}Tr(ad(u^i)\,ad(u^j))
=\sum_{j=1}^{\dim\u}-X_{\alpha\beta}^{j}y_{s(i)}\delta_{ij}
=-X_{\alpha\beta}^{i}y_{s(i)}, $$ with the r.h.s.\ of \rf{Formen}
$$Tr(ad([u^i,p_{\alpha}])\,ad(
p_{\beta}))=\sum_{\gamma=1}^{n}M^i_{\gamma\alpha}
Tr(ad(p_{\gamma})ad(p_{\beta}))=
\sum_{\gamma=1}^{n}M^i_{\gamma\alpha}\delta_{\gamma\beta}
=M^i_{\beta\alpha}.$$
 It follows that
$$\qquad\qquad\qquad\qquad\qquad \qquad
 X_{\alpha\beta}^{i}=-\frac{1}{y_{s(i)}}M^i_{\beta\alpha}=
\frac{1}{y_{s(i)}}M^i_{\alpha\beta}~.
\qquad\; \qed$$

\begin{theorem}[Goddard, Nahm, Olive]  Let $U$ be a compact Lie group
 with a faithful $n$-dimensional orthogonal representation on $\p$.
 Consider the Lie algebra $\u$ of $U$ as a subalgebra of $so(n)$ with
 the inclusion of $\u$ in $so(n)$ induced by the representation on $\p$.
Let $c_K$ denote the central element of the
coset Virasoro algebra on a level one representation
space of $\tilde\Loop (so(n))$.\\ 
Then  $c_K$
 vanishes if and only if $\g\equiv\u\oplus\p$
 carries the structure of a $\Z/2\Z$-graded Lie algebra with 
$\g_0=\u$ and $\g_1=\p$, which is related  to the given orthogonal 
$\u$-module-structure on $\u\oplus\p$ as in lemma \ref{lprod}.
\end{theorem}

\proof The conditions of the theorem, in view of lemma \ref{lprod},
leave no freedom in
defining the Lie product $[.\,,.]_{\g},$ if it exists: 
This is because these conditions already uniquely determine a
bilinear antisymmetric product on $\g=\u\oplus\p$, as follows:
 \br
[u^i,u^j]_{\g}&=&[u^i,u^j]_{\u}\nn\cr
[u^i,p_\alpha]_{\g} &=& \sum_{\gamma=1}^n
M_{\gamma\alpha}^i p_{\gamma} = -[p_\alpha,u^i]_{\g}\nn\cr
[p_{\alpha},p_{\beta}]_{\g} &=&\sum_{i=1}^{\dim\u}
\frac{1}{y_{s(i)}}M^i_{\alpha\beta} u^i 
\er 
(cf.\, lemma \ref{lprod}).
  Therefore, the
proof of the theorem boils down to showing, that $c_K$ vanishes if and
only if this bilinear antisymmetric product on $\g$
obeys the Jacobi identity. 

It can easily be shown from the assumptions, that the Jacobi 
identity holds for any three elements, if at least one of them 
lies in $\u$.
Therefore, $\g$ is a Lie algebra iff the Jacobi identity holds on
its $\p$ part, i.e.\ iff
\be
\label{jid}0=
[p_{\alpha},[p_{\beta},p_{\gamma}]]+[p_{\gamma},[p_{\alpha},p_{\beta}]]+
[p_{\beta},[p_{\gamma},p_{\alpha}]]=\sum_{\delta=1}^n
J_{\alpha\beta\gamma\delta}\,
p_\delta~ \qquad\forall\, \alpha,\beta,\gamma,\end{equation}  
where we used the equation
$$[p_{\alpha},[p_{\beta},p_{\gamma}]]=
\sum_{\delta=1}^n \left( \sum_{i=1}^{\dim\u} 
\frac{M^i_{\gamma\beta}M^i_{\delta\alpha}}{y_{s(i)}}\right) p_{\delta}~,$$ 
which follows from lemma \ref{lprod}. 

Thus, the Jacobi identity \rf{jid} is equivalent to the
condition
  \be
\label{jacobiGNO}
 J_{\alpha\beta\gamma\delta}=0 \qquad\forall\,
\alpha,\beta,\gamma,\delta. \end{equation}

But since the $M^i$ are real, the $J_{\alpha\beta\gamma\delta}$
are also real. Therefore, each individual 
$J_{\alpha\beta\gamma\delta}$ vanishes if and only if the sum of the
squares $ (J_{\alpha\beta\gamma\delta})^2$ on the l.h.s.\ of \rf{g3l}
vanishes. But since by lemma \ref{lJ} this sum is equal to 
$\frac{6}{n}\,c_{\u}\,c_K$ and  $c_{\u}$
is always positive, it follows that \rf{jacobiGNO} is equivalent to
the vanishing of $c_K$:
\be\label{Jc}
c_K=0,\quad\quad \mbox{iff }\quad \quad J_{\alpha\beta\gamma\delta}=0
\qquad\forall\,\alpha,\beta,\gamma,\delta.
\ee
\qed

The equivalence \rf{Jc} was shown in \cite{GO1} 
using the  quark model construction.
This result and the observation that \rf{jacobiGNO} is equivalent to a
Jacobi identity led to the symmetric space
theorem \cite{GNO}.

\bigskip  I would like to thank Prof.\ P.\ Slodowy for awaking my
interest in this subject and for his encouragement and advice.   


\begin{thebibliography}{x}

\bibitem[1]{GNO} P.\ Goddard, W.\ Nahm and D.\ Olive,
Phys.\ Lett.\ 160B, 111-116 (1985)

\bibitem[2]{Witt} E.\ Witt,
Abhandlungen des mathematischen Seminares
der Universit\"at Hamburg 14, 289-322 (1941)

\bibitem[3]{Hum} J.E.\ Humphreys, {\it Introduction to Lie Algebras and
 Representation Theory} (Springer, New York-Heidelberg-Berlin, 1972). 

\bibitem[4]{Kac} V.G.\ Kac, {\it Infinite dimensional Lie algebras
3.Ed.} (Cambridge University Press, 1990). 

\bibitem[5]{GO1} P.\ Goddard und D.\ Olive, Nuclear Physics B257, 
226-252 (1985) 

\bibitem[6]{dab} C.\ Daboul, "Endliche Reduktion bei Darstellungen von
affinen Kac-Moody-Algebren", Hamburger Beitr\"age zur Mathematik
aus dem Mathematischen Seminar Heft 30, Universit\"at Hamburg (1993)

\bibitem[7]{Helg} S.\ Helgason, {\it Differential Geometry, Lie Groups,
 and Symmetric Spaces}  (Academic Press, New York, 1978).

\bibitem[8]{BtD} T.\ Br\"ocker und T.\ tom Dieck, {\it Representations
of  Compact Lie Groups} (Springer, New York-Heidelberg-Berlin, 1985).

\end{thebibliography}
 \end{document}